\documentstyle[11pt,aaspp]{article}

\lefthead{Lilly et al.}
\righthead{Luminosity density to z $\sim$ 1}

\tighten

\begin{document}

\title{The Canada-France Redshift Survey XIII:\\
The luminosity density and star-formation history 
of the Universe to z $\sim$ 1}

\author{S. J. Lilly\altaffilmark{1}}
\affil{Department of Astronomy, University of Toronto, 60 St George Street,
Toronto, ON M5S 3H8, Canada }

\author{O. Le F\`evre\altaffilmark{1}, F. Hammer\altaffilmark{1}} 
\affil{DAEC, Observatoire de Paris-Meudon, Place Jules-Jannsen, 
92195 Meudon, France} 

\author{David Crampton\altaffilmark{1}} 
\affil{Dominion Astrophysical Observatory, West Saanich Road,
       Victoria, BC V8X 4M6, Canada} 
 
\altaffiltext{1} {Visiting Astronomer, Canada-France-Hawaii Telescope, which
is operated by the National Research Council of Canada,
the Centre National de la Recherche Scientifique of France and the University of Hawaii}

\begin{abstract}

The comoving luminosity density of the Universe, ${\cal L}(\lambda)$
is estimated from the CFRS faint galaxy sample in three wavebands 
(2800 $\AA$, 4400 $\AA$ and 1 \micron) over the redshift 
range $0 < z < 1$.
In all three
wavebands, ${\cal L}$ increases markedly with redshift.
For a ($q_0 = 0.5, \Omega = 1.0)$ cosmological model, the
comoving luminosity density increases 
as $(1+z)^{2.1 \pm 0.5}$ at 1 \micron,
as $(1+z)^{2.7 \pm 0.5}$ at 4400 $\AA$ and
as $(1+z)^{3.9 \pm 0.75}$ at 2800 $\AA$, these
exponents being reduced by 0.43 and 1.12
for (0.05,0.1) and (-0.85,0.1) cosmological models respectively.
The ${\cal L}(\lambda)-\tau$ relation 
can be
reasonably well modelled by
an actively evolving stellar population with 
a Salpeter initial mass function (IMF) extending to
125 M$_{\sun}$, and a star-formation
rate declining as $\tau^{-2.5}$ with a turn-on of star-formation
at early epochs.
A Scalo (1986) IMF extending to the same mass limit
produces too many long-lived low mass stars. 
This rapid evolution of the star-formation rate and
comoving luminosity density of the Universe
is in good agreement with the conclusions of
Pei and Fall (1995) from their analysis of
the evolving metallicity of the Universe.
One consequence of this evolution is
that the {\it physical} luminosity density
at short wavelengths
has probably declined by two orders of magnitude since $z \sim 1$.

\end{abstract}


\section{INTRODUCTION}

Recent deep redshift surveys, such as the Canada-France Redshift Survey
(CFRS), have produced large samples of normal field galaxies at high redshifts
out to $z > 1$.  A great deal can be learnt about the 
evolution of galaxies from the study of
these galaxies, both through the construction of distribution functions
such as the luminosity function and through the detailed study of 
individual galaxies. However, for some purposes, it is of 
interest to study the integrated light from the whole population, i.e. 
the comoving luminosity density, ${\cal L}(\lambda,z)$.
The principal motivation is that this quantity is independent of many of the 
details
of galactic evolution and depends primarily on the global star-formation 
history of the Universe and the, possibly epoch-dependent, initial 
mass function (IMF) of the stars.  
In particular, ${\cal L}(\lambda,z)$ should be independent of
the merging history of 
individual galaxies, and independent of the uncertainties concerning the comoving
densities and timescales of rapidly evolving galaxies (i.e. whether a given object
represents a short-lived evolutionary phase
occuring in many galaxies or a 
longer duration phenomenon occuring in just a few).

Analysis of ${\cal L}(\lambda,z)$ thus offers the prospect of
determining the global star-formation history of the Universe. 
It is reasonable to hope that the star-formation rate 
averaged over the entire galaxy population in the Universe
might follow a relatively simple dependence on cosmic epoch,
even if individual galaxies have more stochastic
evolutionary histories.
Other cases where the integrated comoving
luminosity density of the Universe
will be of interest include (a) the calculation of the change in the 
average metallicity
in the Universe as sampled by quasar absorption lines (see e.g.
Pei and Fall 1995); 
(b) the calculation of the expected rate of supernovae; and
(c) the calculation of the closure mass-to-light
ratio.

In this {\it Letter}, we construct
the comoving luminosity density of the Universe at three wavelengths
(2800
\AA, 4400 \AA, and 1 $\micron$), over the redshift interval 
$0 < z < 1$, from the
CFRS
galaxy sample.
We consider three representative
cosmological models: ($q_0 = 0.5 ,\Omega_0 = 1$) and
(0.05, 0.1) Friedman models, and a low density zero-curvature model,
(-0.85, 0.1). We take
$H_0 =
50 h_{50}$ kms$^{-1}$Mpc$^{-1}$ in computing volume elements.

\section{CONSTRUCTION OF THE COMOVING LUMINOSITY DENSITY}

\subsection{Estimation of the $0.2 < z < 1.0$ luminosity density from the CFRS}

The CFRS galaxy sample has been
described in detail elsewhere
(Lilly et al. 1995a, Le F\`evre et al. 1995a,
Lilly et al. 1995b,
Hammer et al. 1995, and Crampton et al. 1995; CFRS V).
It
consists of 730 $I$-band selected galaxies ($17.5 < I_{AB} < 22.5$),
of which 591 (i.e. more than 80\%) have secure redshifts in
the range $0 < z < 1.3$, with a
median $<z> \sim 0.56$. All objects have $V$ and $I$ photometry, and 
most have also been observed in $B$ and
$K$, allowing
the spectral energy distribution to be defined
over a long wavelength baseline.
Scientific analyses relevant to the subject of this
{\it Letter} include analysis 
of the luminosity function (Lilly et al. 1995c; CFRS VI), 
spatial correlation function
(Le Fe\`vre et al. 1996) and galaxy morphologies
(Schade et al 1995).

The contribution to the comoving luminosity density from directly observed sources may
be easily estimated using the $V_{max}$ formalism used to construct the 
luminosity function (see CFRS VI). 
 
\[{\cal L}(\lambda) = \sum_{i} \frac{L_i(\lambda)}{V_{max,i}}\]

with 
$V_{max,i}$ for each galaxy, $i$, defined and computed as in CFRS VI.
The luminosity density has been computed at
rest-4400 $\AA$, rest-2800 $\AA$ and rest-1 \micron.
The estimates of
$L_i(\lambda)$ are thus
{\it interpolations} of
our available $BVIK$ photometry, except for the rest-2800 $\AA$
estimate which is a modest extrapolation for $z < 0.5$.
The spectral energy distributions (SEDs) for individual galaxies 
are 
interpolated from the four SEDs given 
by Coleman et al. (1980; CWW) so as to match the 
observed colors
of each galaxy.  

One source of uncertainty arises from our failure to obtain secure
redshift 
identifications for
19\% of the galaxies. The likely nature of these galaxies
is discussed in CFRS V.
In computing the luminosity density, we have assigned all of the 
unidentified galaxies a redshift based on their photometric properties.
This procedure
introduces
a maximum 10\% uncertainty in $\cal L$ over the $0.2 < z < 1.0$ range.
The uncertainty at the extremes of the CFRS $N(z)$,
at $z < 0.2$ and $z > 1.0$, is obviously larger, and 
we do not present the
luminosity density in these redshift regimes. 
Our uncertainty in ${\cal L}(\lambda)$ is estimated from
the sum in quadrature of the bootstrap error obtained by resampling the sample
(see CFRS VI), the 10\% error arising from the spectroscopic incompleteness,
plus a Poisson error of $(2.5/N)^{0.5}$ which reflects the fact that 
galaxies are clustered in redshift space (see CFRS VI and CFRS VIII). 
We refer to the luminosity densities derived in this way as the
``directly observed'' luminosity densities. They are listed in Table
1 in units of $h_{50}$ WHz$^{-1}$Mpc$^{-3}$. For reference,
the Sun has 
$L(4400) = 3.4 \times 10^{11}$ WHz$^{-1}$ and
one
present-day L* galaxy ($M_{AB}(B) = -21.0 + 5 {\rm log} h_{50}$)
per Mpc$^{-3}$ produces a luminosity density of $1.14 \times 10^{22}
h_{50}^{-2}$ WHz$^{-1}$Mpc$^{-3}$.

A second major uncertainty is the contribution from galaxies whose
individual luminosities place them below the magnitude limit of the
CFRS sample. We have approached this problem by fitting
the rest-$B$ luminosity functions of blue and red galaxies, as
derived in different redshift intervals in CFRS VI,
and integrating these over all luminosities
to produce the luminosity density in the 
rest-frame $B$-band. 
The luminosity function of red and blue populations 
are assumed to 
have constant shape ($\alpha = -1.3$ for blue and $\alpha = -0.5$ for red)
but a normalization
in both $M*$ and $\phi$* that varies with epoch. 
As seen in Table 1, the increase in this ``LF-estimated'' 
luminosity density over that
``directly-observed''
in the CFRS galaxies is modest at $z < 0.75$, but quite large (a
factor of about two) for $0.75 < z < 1.0$.
A formal uncertainty in this procedure
was estimated by considering the range of ($M*,\phi*$) values that
gave acceptable fits to the observed luminosity function, but this estimate
is unrealistically small,
so a $1\sigma$ uncertainty of 33\% of the additional flux was adopted
(i.e. providing a 
$\pm 1\sigma$ range of a factor of two in the additional flux).
Luminosity densities at the other wavelengths were then produced by applying
luminosity-weighted average colors, $(2800-4400)_{AB}$ and $(4400-10000)_{AB}$, 
derived from the ``directly-observed'' CFRS galaxies.

\subsection{The local luminosity density}

There have been several recent determination of the local luminosity function
based on surveys
of several thousand galaxies.
Converting the photometric systems as best we can 
(some are quite poorly defined) and integrating the luminosity functions
we obtain values of log({\cal L}(4400)) (with the same units as above)
of 19.30 (Loveday et al. 1992),
19.40 (da Costa et al. 1994), 19.55 (Marzke 1994a) and 19.22 (Marzke 1994b).
The $r$-selected luminosity function of Lin et al. (1995) gives 
19.24 if the average $(B-r) \sim 1.0$.  A simple average of all of the above
gives $19.34 \pm 0.06$ and
the Loveday (1992) value, 19.30, was finally adopted, with an
uncertainty of 0.1 in the log.
The local luminosity densities at 2800 $\AA$ and 1 $\micron$ have 
been estimated 
by applying luminosity weighted colours to the $B$-band luminosity density.
These colors were estimated using the CWW SEDs
from the Marzke et al. (1994b) morphological
type-dependent luminosity function 
and from the Metcalfe et al. (1991)
(B-V)-dependent luminosity function.
These 
two estimates agree to within 0.15 in the (4400-10000) color
and to within 0.3 in the (2800-4400) color and 
We assign uncertainties of 0.20 and 0.10 in the log to the estimates of the 
local luminosity
density at 2800 $\AA$ and 1 \micron (see Table 1).

\section{THE EVOLUTION OF THE LUMINOSITY DENSITY}

\subsection{Parameterization as $(1+z)^{\alpha}$}

As shown in Figure 1, the ``LF-estimated'' luminosity densities
are well-represented by power-laws in $(1+z)$.
If we write
${\cal L}(\lambda,z) = {\cal L}_0(\lambda) (1+z)^{\alpha(\lambda)}$ then we find
$\alpha(2800) = 3.90 \pm 0.75$,
$\alpha(4400) = 2.72 \pm 0.5 $, and
$\alpha(10000) = 2.11 \pm 0.5$ for the (0.5,1.0)
cosmology as plotted on the Figure.
Changing
to
$q_0 = 0.0$
introduces an offset exactly equal to 0.5 log$(1+z)$. The lowest panel in Figure 1 
shows that in the two alternative cosmological models considered,
(0.05,0.1) and
(-0.85,0.1),
the slopes would be decreased 
by approximately 0.43 and 1.12 respectively.

\subsection{Parameterization with epoch}

In terms of stellar populations, it is
of interest to plot these changing luminosity densities
as a function of cosmic epoch. In view of the uncertainty
in both $H_0$ and $q_0$, and since we wish to compare the evolving luminosity densities 
with stellar population models, a reasonable approach is to 
normalize the present age of the 
Universe to 15 Gyrs, as representative of the age of the oldest stellar 
populations.
This therefore implies values of $H_0$ of
45, 60 and 85 kms$^{-1}$Mpc$^{-1}$ for the (0.5,1.0), (0.05,0.1) and 
(-0.85,1.0) cosmologies respectively. 

The upper panel of Figure 2 shows the comoving luminosity densities
as a function of epoch. 
For simplicity,
the $L(\lambda)$ have been transcribed from
Figure 1 {\it without} correction for
the different {\it volume elements}
implied by these different values of
$H_0$.
Straight lines have been fit to the data for each cosmological model, 
and the gradients,
$d$(log${\cal L}(\lambda) / dt$, are listed in Table 2.
The variation of ${\cal L}(\lambda)$ is almost independent of $q_0$,
especially for the zero-$\lambda$ models and the longer wavelengths
(4400 $\AA$ and 1 $\micron$).
The luminosity density declines faster than expected for
the purely passive evolution of an old stellar population, by which we mean
the evolution (due to main-sequence burn-down) of a stellar population
that has no continuing star-formation 
after an initial burst.
In their 
comparison of models, Charlot, Worthey and Bressan (1996) find, 
with a Salpeter IMF 
and solar metallicity, 
$d$(log${\cal L}(4400) / dt = 0.042 \pm 0.008$ and
$d$(log${\cal L}(22000) / dt = 0.029 \pm 0.006$ for ages between 5-17 Gyr,
about a half of what is observed.

\subsection {Towards a global history of star-formation?}

An interesting question arises as to
whether a simple model, defined by a time-dependent
star-formation rate and a time-independent initial mass function, can,
ignoring the 
effects of reddening by dust, reproduce the form of Figure 2. 
A preliminary investigation
suggests that the answer is probably ``yes'',
provided that the IMF is reasonably 
rich in massive stars.  
In the two lower panels of Figure 2 two sets
of simple models, generated
from the GISSEL library (Bruzual and Charlot 1992), are compared with 
our data.
The middle panel shows model stellar populations
with the Salpeter power-law IMF 
(with $x$ = 1.35) with 
0.1 M$_{\sun} < M < 125 $M$_{\sun}$
that reproduce the ${\cal L}(2800)$ over the
redshift range $0 < z < 1$.  Since the light at 2800 $\AA$ is
dominated by very young stars with this IMF, 
the models are required
to have a time-dependent
star-formation rate proportional to $\tau^{-2.5}$ over the 
relevant range of epochs (or slightly steeper, $\tau^{-3}$, for
the (0.05,0.1) cosmology).
The four models shown in the middle panel
all share this $\tau ^{-2.5}$ star formation history but have a
turn-on of star-formation 2,3,4 and 5 Gyr after the Big Bang.
A model starting 2 to 3 Gyr after the Big Bang
would clearly give a reasonable representation
of the data, given the uncertainties in both data and models.
The gradients in ${\cal L}(\lambda,\tau)$ and the 15 Gyr colors of the
3 Gyr model
are listed in Table 2.

Initial mass functions which are less rich in massive
stars do less well. In the lower panel of Figure 2, the predictions of similar
models based on the Scalo (1986) IMF,
which has almost an order of magnitude
fewer high mass stars relative to solar mass stars, are shown. 
To match the
ultraviolet luminosity density, the star-formation rate is
set to $\tau^{-2.8}$, but the models are otherwise similar to those
shown the the middle panel. These models
all produce too much long wavelength light (i.e. at 4400 $\AA$ and at 1 \micron)
by the present epoch and/or
do not decline in brightness at the longer wavelengths
fast enough because of the 
large number of solar mass stars produced in the star-formation activity
required to yield the high ultraviolet ${\cal L}(2800)$ at
high redshifts. The adoption of the
(-0.85,0.1) cosmology would alleviate this problem by reducing 
the required evolutionary gradient at 1 \micron.
However, 
uncertainties in the stellar models (e.g. see Charlot, Worthey and Bressan
1996) are still large enough to suggest
that these conclusions concerning the IMF should be treated with
caution.

The above analysis has ignored the effects of reddening due to dust
(as well as other metallicity related effects in the models). 
If the effect of dust was
independent of epoch, then it would simply
shift the long wavelength
curves of each model to higher values (since the models are
normalized
to the observed
${\cal L}(2800)$), thus exacerbating the problem with the Scalo IMF.

\section{DISCUSSION AND SUMMARY}

Analysis of the CFRS galaxy sample
indicates that 
the observed luminosity density of the Universe in the ultraviolet, optical
and near-infrared wavebands increases markedly with redshift
over $0 < z < 1$, as
${\cal L} \propto (1+z)^{2.1 \pm 0.5}$ at 1 \micron,
${\cal L} \propto (1+z)^{2.7 \pm 0.5}$ at 4400 $\AA$ and
${\cal L} \propto (1+z)^{3.9 \pm 0.75}$ at 2800 $\AA$, for the
(0.5,1.0) cosmology. The
exponents would be reduced by 0.43 and 1.12
for (0.05,0.1) and (-0.85,0.1) cosmological models respectively.
If an
IMF rich in massive stars is assumed (as seems to be required) 
and if the effects of dust are ignored,
then the ultraviolet
luminosity density translates more or less 
directly to the star-formation rate,
implying a 
rapid decline in the overall star-formation rate since $z \sim 1$. 

At $z \sim 1$, the global star-formation rate
was a factor of 15 higher for (0.5,1.0),
11 times higher for (0.05,0.1) and 7 times
higher for (-0.85,0.1), with an uncertainty
in each case of about 0.22 in the log.
These large increases are in remarkably good agreement with the 
quite independent estimates of Pei and Fall (1995),
a factor of
20 increase for (0.5,1.0) and of 10 for (0.0,0.0),
based on a careful modelling of the
change in mean metallicity of the
Universe.

The {\it physical} luminosity density at 2800 A has
evidently declined as $(1+z)^{6.7 \pm 0.75}$ for a (0.5,1.0)
cosmological model, i.e. by 
a factor of
60-170 since $z \sim 1$. This large factor is reduced by only a factor of
two even
in the extreme (-0.85,0.1) model. 

\acknowledgements

We were initially encouraged to
compute the luminosity density
by Mike Fall, and we have greatly benefitted from several
subsequent discussions with him and with
Ray Carlberg. 
SJL's 
research is supported by the NSERC of Canada 
and the CFRS project has been facilitated by a 
travel grant from NATO.

\clearpage

\begin{figure}

\caption{
The comoving luminosity density (WHz$^{-1}$Mpc$^{-3}$) of the Universe 
in three wavebands.  The small
symbols at $z > 0.2$ are the ``directly-observed'' luminosity 
density, the larger symbols with error bars are the ``LF-estimated''
luminsity density (see text). 
The population of galaxies redder than Sbc are shown as squares,
those bluer than Sbc as open circles and
that of all galaxies as solid circles.  The solid line shows the best-fit power-laws
for the ``LF-extimated''
luminosity density for ``all'' galaxies
in each waveband. The luminosity densities are
calculated for the (0.5,1.0) cosmological model. The
panel at bottom shows the 
roughly linear offset that should be applied for the (0.05,0.1) 
and (-0.85,0.1) cosmological models.
}

\caption{
The comoving ``LF-estimated'' luminosity density for ``all'' galaxies from Figure 1,
plotted as a function of epoch. The epoch is normalized so that the present epoch is 15 Gyr,
irrespective of the cosmological model (see text).
In the upper panel, straight lines have been fit to the data 
for each cosmological model.
The solid circles and solid line represent the (0.5,1.0) cosmological model, 
open squares and dashed line, the
(0.05,0.1) model, and the triangles and dotted line, the (-0.85,0.1) model.  
The dashed curve in the upper panel labelled CWB is the passively
evolving model of Charlot et al. (1996).
The same
data points are reproduced on the two lower panels. The middle
panel shows stellar population models with a Salpeter initial mass function,
the bottom panel shows models with a Scalo (1986)
initial mass function, in both 
cases the IMF extends over the range $0.1 M\sun < M < 125 M\sun$. 
In each case, four models are shown which have the same star-formation history
($\propto \tau ^ {-2.5}$ with the Salpeter IMF, $\propto \tau ^ {-2.8}$
with the Scalo IMF) but different turn-ons at 2, 3, 4, and 5 Gyr.}

\end{figure}

\begin{references}
     
\reference Charlot, S., Bruzual, G.A., 1993, ApJ, 405, 538
\reference Charlot, S., Worthey, G., Bressan, P., 1986, ApJ, {\it in press}.
\reference Coleman, G.D., Wu, C.C., Weedman, D.W., 1980, ApJ(Supp), 43, 393 (CWW)
\reference Crampton, D., Le F\`evre, O., Lilly, S.J., Hammer, F., 1995, ApJ, 
455, 96 (CFRS V)
\reference Da Costa, L.N., Geller, M.J., Pellegrini, P.S.,
Latham, D.W., Fairall, A.P., Marzke, R.O., Wllimer, C.N.A., Huchra, J.P., Calderon, J.H.
Ramella, M., Kurtz, M.J.; 1994, ApJ(Letter), 424, L1.
\reference Hammer, F.,  Crampton. D., Le F\`evre, O., Lilly, S.J., 1995a, ApJ,
455, 88 
\reference Le F\`evre, O., Crampton. D., Lilly, S.J., Hammer, F., Tresse, L., 1995, ApJ, 455, 60
\reference Le F\`evre, O., Hudon, D., Lilly, S.J., Crampton. D., Hammer, F., 1996, ApJ, in press.
\reference Lilly, S.J., Le F\`evre, O., Crampton. D., Hammer, F.,Tresse, L., 1995a, ApJ, 455, 50
\reference Lilly, S.J., Hammer, F., Le F\`evre, O., Crampton. D., 1995b, ApJ, 455, 75
\reference Lilly, S.J., Tresse, L., Hammer, F., Crampton, D., Le Fe\`vre, O.,;
1995c, ApJ, 455, 108 (CFRS VI)
\reference Lin, H., Kirshner, R.P., Shectman, S.A., Landry, S.D., 
Oemler, A., Tucker, D.L., Schechter, P.L.; 1996, ApJ, submitted.
\reference Loveday, J., Peterson, B.A., Efstathiou, G., Maddox, S.J., 1992, ApJ, 390, 338
\reference Marzke, R.O., Geller, M.J., Huchra, J.P., Corwin, H., 1994a, ApJ 428, 43
\reference Marzke, R.O., Geller, M.J., Huchra, J.P., Corwin, H., 1994b, AJ 108, 437
\reference Metcalfe, N., Shanks, T., Fong, R., Jones, L.R, 1991, MNRAS, 249, 498
\reference Pei, Y.C., Fall, S.M., 1995, ApJ, 454, 69
\reference Schade, D.J., Lilly, S.J., Crampton, D., Hammer, F., Le F\`evre, O., Tresse, L., 
1995, ApJLett, 451, L1


\end{references}
\end{document}